\documentclass[twocolumn,showpacs,superscriptaddress,amsmath,amssymb]{revtex4}

\usepackage{graphicx}
\usepackage{dcolumn}
\usepackage{bm}

\begin{document}

\title{
Coherent and Incoherent States of Electron-doped SrTiO$_3$}

\author{Yukiaki~Ishida}
\author{Ritsuko~Eguchi}
\author{Masaharu~Matsunami}
\author{Koji~Horiba}
\author{Munetaka~Taguchi}
\author{Ashish~Chainani}
\affiliation{RIKEN SPring-8 Center, Sayo, Sayo, Hyogo 679-5148, 
Japan}

\author{Yasunori~Senba}
\author{Haruhiko~Ohashi}
\affiliation{JASRI/SPring-8, Sayo, Sayo, Hyogo 679-5198, 
Japan}

\author{Hiromichi~Ohta}
\affiliation{Graduate School of Engineering, Nagoya University, 
Furo, Chikusa, Nagoya, Aichi 464-8603, Japan} 

\author{Shik~Shin}
\affiliation{RIKEN SPring-8 Center, Sayo, Sayo, Hyogo 679-5148, 
Japan}
\affiliation{ISSP, University of Tokyo, Kashiwa-no-ha, Kashiwa, 
Chiba 277-8561, Japan}

\date{\today}

\begin{abstract}
Resonant photoemission at the Ti 2$p$ and O 1$s$ edges on 
a Nb-doped SrTiO$_3$ thin film 
revealed that the coherent state (CS) at the Fermi level ($E_{\rm F}$) 
had mainly Ti 3$d$ character 
whereas the incoherent in-gap state (IGS)
positioned $\sim$1.5\,eV below $E_{\rm F}$ 
had mixed character of Ti 3$d$ and O 2$p$ states. 
This indicates that 
the IGS is formed by a spectral-weight transfer from the 
CS and subsequent spectral-weight redistribution 
through $d$-$p$ hybridization. 
We discuss the evolution of the excitation spectrum 
with 3$d$ band filling and rationalize the IGS 
through a mechanism similar to that proposed by 
Haldane and Anderson. 
\end{abstract}

\maketitle

SrTiO$_3$ (STO) is a perovskite-type oxide semiconductor 
with a band gap of 3.2\,eV \cite{Cardona} and 
is considered to be one of the promising device materials 
in ``oxide electronics."
A variety of transition-metal oxides can be grown on STO 
with atomical flatness to show novel phenomena such as 
the high-mobility and magnetic interface of 
LaAlO$_3$/STO \cite{HighMob, Brinkman}
and state-of-the-art thermoelectric performance of STO:Nb/STO 
superlattice \cite{Ohta_NM}. 
Additionally, 
carrier (electron) concentration of STO is controllable 
through 
substitutional doping 
or through a field effect \cite{Takagi}, 
as performed in conventional semiconductor-device materials.

\begin{figure}[htb]
\begin{center}
\includegraphics[width=8.5cm]{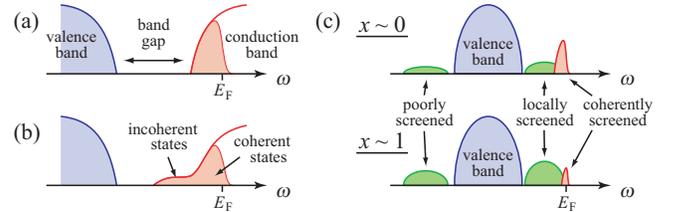}
\caption{\label{fig0} 
Schematic electronic structures near $E_{\rm F}$ 
within band theory (a), when correlation effects 
are included for the conduction electrons (b), and 
within a model including various screening channels (c). }
\end{center}
\end{figure}

While band theory may be a starting point to understand the 
transport and magnetic properties of lightly-doped STOs \cite{Tokura}, 
their electronic structures near the Fermi level ($E_{\rm F}$) 
are far from being doped band insulators: 
photoemission spectroscopy (PES) studies 
\cite{Fujimori_LSTO, Yoshida_EPL, Higuchi_STO_donor, Aiura_ARPES, DDSarma} 
have revealed 
incoherent states forming in the band gap of 
STO (in-gap states: IGSs)
instead of a rigid shift of the bands with increasing the conduction-band 
filling $x$ 
as schematically shown in Fig.\,\ref{fig0}(a). 
Since the system shows a metal-insulator transition 
with $x$\,$\to$\,1 and the end member is a correlated $d^1$ insulator 
such as LaTiO$_3$ \cite{Tokura} (a Mott-Hubbard-type insulator in the 
Zaanen-Sawatzky-Allen scheme \cite{ZSA_diagram}), 
the IGSs in the metallic phase may be viewed as 
precursurs of the ``lower-Hubbard band" of the $d^1$ insulators 
\cite{Fujimori_92}. 
However, a Hubbard model, which produces incoherent states 
through electron correlation as shown in Fig.\,\ref{fig0}(b), 
is insufficient to explain the IGSs \cite{Fujimori_LSTO, Robey, Morikawa} 
since they are observed already at $x\sim$ 0.05 
\cite{Aiura_ARPES} where electron correlation is expected to be small 
\cite{Tokura}. 
The origin of the IGSs at $x\sim$ 0 
was attributed to 
chemical disorder \cite{DDSarma}, 
polaron effect \cite{Fujimori_Polaron}, 
or partly to donor levels \cite{Higuchi_STO_donor}, 
but no consensus has been reached yet (see p.\,1171 of \cite{RMP}).

In this Letter, we investigate the origin of the IGSs 
of lightly-electron-doped STO using soft-x-ray resonant PES, 
which is a convenient tool to obtain the partial density of states 
(PDOS) in the valence-band spectra \cite{Davis}. 
By performing resonant PES at the Ti 2$p$ and O 1$s$ edges, 
we find not only Ti 3$d$ character but also O 2$p$ character 
in the IGSs, which is direct evidence for $d$-$p$ hybridization playing an 
important role for the existence of the IGSs, 
whereas the coherent states at $E_{\rm F}$ has mainly Ti 3$d$ character. 
In contrast to the picture that 
the IGSs and the coherent states are resulting from 
correlations within the conduction electrons 
as schematically shown in Fig.\,\ref{fig0}(b), we assign them 
to locally-screened and non-locally-screened final states, respectively, 
as shown in Fig.\,\ref{fig0}(c). 
The IGSs are considered qualitatively equivalent 
to the multiple charge states 
of transition-metal impurities in semiconductors 
accumulating in the middle of the band gap with 
increasing hybridization as proposed by 
Haldane and Anderson \cite{HaldaneAnderson}. 

A 100-nm thick STO:Nb film was grown epitaxially on a 
(100) face of insulating LaAlO$_3$ by a pulsed laser deposition 
method \cite{Ohta_STON_film}. 
The carrier concentration was estimated to be 
$\sim$1$\times$10$^{21}$\,cm$^{-3}$ or $x\sim$ 0.06 (see below). 
Soft-x-ray 
absorption (XAS) and PES measurements were performed at 50\,K 
at undulator beamline BL17SU of SPring-8 equipped with 
a Gammadata-Scienta SES2002 analyzer \cite{Ohashi}. 
The incident light was circularly polarized. 
The PES spectra were recorded at $\sim$250-meV energy resolution 
and the binding energies ($E_{\rm B}$'s) were referenced 
to $E_{\rm F}$ of gold in contact with the sample and the analyzer. 
An STO film grown on an STO:Nb/LaAlO$_3$ was also measured. 
The sample surface was reasonably clean without any surface treatment, 
as described below. 

\begin{figure}[htb]
\begin{center}
\includegraphics[width=7.5cm]{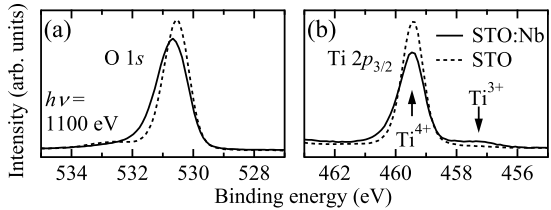}
\caption{\label{fig1} O 1$s$ (a) and Ti 2$p$$_{3/2}$ (b) core-level spectra. }
\end{center}
\end{figure}

\begin{figure}[htb]
\begin{center}
\includegraphics[width=8.7 cm]{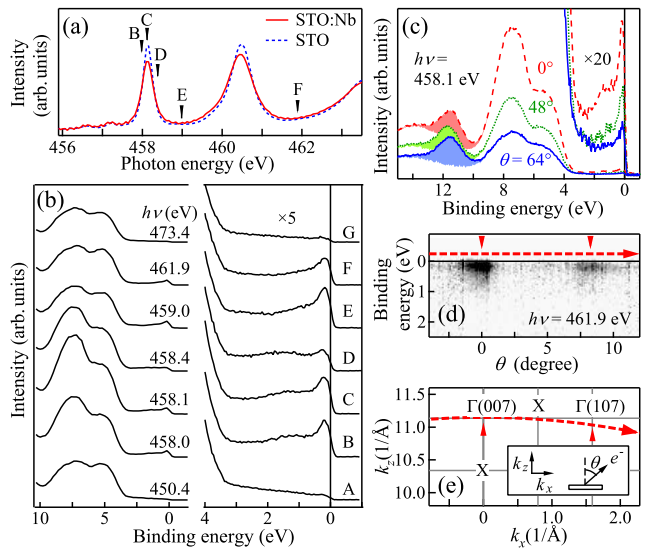}
\caption{\label{fig2} 
Resonant PES at the Ti 2$p$$_{3/2}$ absorption edge. 
(a) Ti 2$p$$_{3/2}$ XAS of STO:Nb and STO. 
(b) Valence-band spectra of STO:Nb at the Ti 2$p$$_{3/2}$ absorption edge. 
The intensities were normalized to the photon flux (post-focusing 
mirror current). 
The labels on the spectra (B to F) correspond to photon energies 
indicated in (a). 
(c) Take-off-angle dependence of the valence-band spectra taken 
in the angle-integrated mode at $h\nu$ = 458.1\,eV. 
Spectra were normalized to the $\sim$11-eV peak area (colored area.) 
(d) Dispersion recorded in the angle-resolved mode 
along a trace in the momentum space shown by a dotted line in panel (e). 
Black/white corresponds to 
high/low intensity. Triangles indicate electron pockets detected 
around the $\Gamma$ points. 
Inset in (e) defines $\theta$ and coordinate of the momentum vector. }
\end{center}
\end{figure}

Figure \ref{fig1} shows core-level spectra of STO:Nb and STO. 
The O 1$s$ spectrum of STO:Nb was a single peak at 
$E_{\rm B} = 530.7$\,eV 
($\equiv E_{1s}$) 
with an asymmetry 
due to metallic screening. 
A hump feature at $E_{\rm B} \sim 533$\,eV was negligible, 
indicating that the 
sample surface was fairly clean. 
The Ti 2$p$$_{3/2}$ spectrum of STO:Nb showed a 
weak Ti$^{3+}$ peak at $\sim$2\,eV below the main Ti$^{4+}$ peak 
at $E_{\rm B} = 459.5$\,eV. 
From the intensity ratio of Ti$^{3+}$/Ti$^{4+}$ peaks 
\cite{Morikawa, Robey}, we deduced $x$ = 0.06$\pm$0.02. 
This value was consistent with $x<0.1$ estimated from the 
Ti 2$p$ XAS line-shape broadening [Fig.\,\ref{fig2}(a)], 
which is a measure of the content of 
Ti$^{3+}$ component \cite{Abbate_2pXAS}. 
Since STO:Nb is a paramagnetic metal, we interpret that 
the valence of Ti is fluctuating between 3+ and 4+ 
with a time scale longer than that characteristic of a photoemission 
process ($\sim$10$^{-16}$\,s) \cite{FastProbe}, and PES provides 
a snapshot of this fluctuation to give a double-peaked Ti 2$p$ spectrum. 
Alternatively, the photoemission initial state of STO:Nb 
is well described as a fraction $x$ of the Ti 
sites have nominally-$d^1$ configuration 
and the rest have nominally-$d^0$ configuration. 
Later, we will use this picture in explaining the 
excitations near $E_{\rm F}$.

Valence-band spectra recorded around the Ti 2$p$$_{3/2}$ absorption 
edge are shown in Fig.\,\ref{fig2}(b). 
At $h\nu = 458.1$\,eV, resonant enhancement was observed in 
the O 2$p$ band region 
[the $\sim$6-eV-wide structure 
centered at $E_{\rm B} = 6.5$\,eV ($\equiv E_{2p}$)] and 
in the near-$E_{\rm F}$ region ($E_{\rm B} \textless 3$\,eV.) 
In the O 2$p$ band region, the higher-$E_{\rm B}$ side showed 
stronger enhancement reflecting Ti 3$d$\,-\,O 2$p$ 
bonding character as described by band theory 
\cite{Higuchi_STO_RPES, Mattheiss, Band_Toussaint}. 
The enhanced structures at $h\nu =$ 458.1\,eV in the near-$E_{\rm F}$ region 
consisted of a peak at $E_{\rm F}$ and a broad structure centered at 
$\sim$1.5\,eV, which are the coherent states and 
the IGSs of doped STO, respectively 
\cite{Fujimori_LSTO, Yoshida_EPL, Higuchi_STO_donor, Aiura_ARPES, DDSarma}. 
We also confirmed that the resonantly enhanced IGSs 
were dominated by bulk character states from the 
take-off-angle ($\theta$) dependence of the spectra at $h\nu = 458.1$\,eV 
as shown in Fig.\,\ref{fig2}(c). 
Note that, if the $\sim$1.5-eV state was due 
to surface states, 
the intensity should scale, with decreasing bulk sensitivity 
(with increasing $\theta$), to that of the 11-eV peak 
which originates from surface adsorbates \cite{Bednorz_PES}. 
This is clearly not the case as seen in Fig.\,\ref{fig2}(c), 
where the intensity at $E_{\rm B}$ $<$ $\sim$2\,eV scales with 
the O 2$p$ band intensity.

In Fig.\,\ref{fig2}(b), one can also see enhancement of the 
coherent state in the spectra recorded at $h\nu=$ 459.0 and 461.9\,eV. 
This is attributed to normal-emission angle-resolved effect, 
that is, an electron pocket around the $\Gamma$ point was 
crossed at $h\nu\sim$ 460\,eV. 
In fact, we confirmed the electron pockets around the $\Gamma$ points 
in an off-normal angle-resolved PES measurement at $h\nu = 461.9$\,eV, 
as shown in Fig.\,\ref{fig2}(d). 
Using an inner potential 12$\pm$1\,eV measured from $E_{\rm F}$ 
\cite{Aiura_ARPES}, 
the $c$-axis lattice parameter was 3.94$\pm$0.02\,\AA, 
which is in good agreement with that derived 
from x-ray diffraction measurements \cite{Ohta_STON_film}. 
The dispersion observed in the angle-resolved PES spectra is a sign 
of good crystallinity of the sample surface.

\begin{figure}[htb]
\begin{center}
\includegraphics[width=8.7cm]{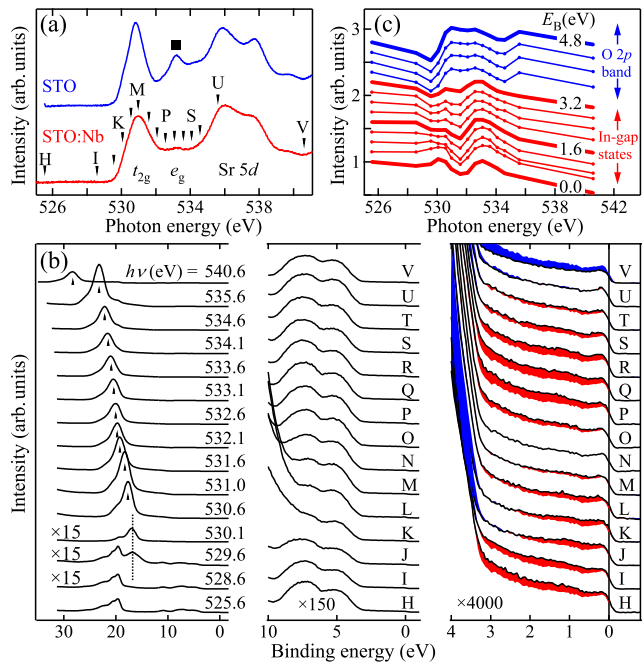}
\caption{\label{fig3} 
Resonant PES at the O 1$s$ absorption edge. 
(a) O 1$s$ XAS of STO:Nb and STO. 
The structures 
at $h\nu = $ 530.9, 533.2, and 534.5\,-\,538.5\,eV 
corresponds to O 1$s$ core-electron excitation into 
O 2$p$ states hybridized with 
Ti 3$d$ $t_{2g}$, Ti 3$d$ $e_{g}$, and Sr 5$d$ states, 
respectively. 
(b) Valence-band spectra at the O 1$s$ absorption edge. 
The labels on the spectra (H to V) correspond 
to the photon energies indicated in (a). 
The triangles in the left panel indicate the 
O $KL_{2,3}L_{2,3}$ Auger peak positions. 
The red 
(light grey) 
and blue 
(dark grey) 
area in the right panel indicate positive and 
negative differences to the 531.6-eV spectrum. 
The structures around 20 and 22\,eV 
are due to Sr 4$p$ and O 2$s$ core levels, respectively. 
(c) CIS spectra. Each spectrum is normalized 
to the intensity at 525.6\,eV and has an arbitrary offset. }
\end{center}
\end{figure}

Figure\,\ref{fig3}(a) shows O 1$s$ XAS of STO:Nb and STO. 
The O 1$s$ XAS line shape of STO is in good agreement with that 
reported by de~Groot {\it et al} \cite{DeGroot_O1sXAS}. The 
overall line shape except for the sharpness of the peak at 
$h\nu\sim$ 533.2\,eV as indicated by a square in Fig.\,\ref{fig3}(a) 
was well reproduced by the 
oxygen-$p$-projected unoccupied 
density of states of STO calculated using a local-density approximation 
\cite{DeGroot_O1sXAS}. 
The sharp peak at $h\nu\sim$ 533.2\,eV 
appearing in the excitation region of O 1$s$\,$\to$\,O 2$p$ states 
hybridized into Ti 3$d$ $e_g$ states 
was attributed to the O 1$s$ core hole in the XAS final state 
\cite{DeGroot_O1sXAS}: 
the electron excited into the $e_g$ band is strongly influenced by the 
O 1$s$ core-hole potential 
through large Ti 3$d$ $e_g$\,-\,O 2$p$ hybridization 
to form a core-excitonic state. 
The peak at $h\nu\sim$ 533.2\,eV 
is supressed and broadened in the spectrum of 
STO:Nb [Fig.\,\ref{fig3}(a)], 
indicating that the life time of the core-excitonic state 
became shorter due to metallic screening. 
Nevertheless, the peak at $h\nu\sim$ 533.2\,eV is still observed in the 
spectrum of STO:Nb, indicating that the localized 
XAS final state is still present. 
Indeed, the resonant enhancement in the near-$E_{\rm F}$ spectra 
occurs at $h\nu\sim$ 533.2\,eV as described below.

Figure\,\ref{fig3}(b) shows valence-band spectra 
recorded around the O 1$s$ absorption edge. 
For $h\nu \ge 530.1$\,eV, 
intense O $KL_{2,3}L_{2,3}$ Auger emission 
\cite{Tjeng_OKRPES, JHPark, Tjernberg, Tezuka_Ti2O3RPES, TiO2_RPES} 
appeared and the peak position ($\equiv E_{KLL}$) shifted to 
higher $E_{\rm B}$'s with increasing $h\nu$'s. 
The O 2$p$ on-site Coulomb interaction was estimated to be 
$U_{pp}\sim 5.5$\,eV using the relationship 
$
U_{pp} = E_{1s} - (h\nu - E_{KLL}) - 2 E_{2p} 
$ 
for $h\nu \ge 534.6$\,eV \cite{JHPark}. 
At $h\nu \leq$ 530.1\,eV, $E_{KLL}$ stays at 
constant $E_{\rm B}$ as 
indicated by a dotted vertical line in Fig.\,\ref{fig3}(b). 
The constant-initial-state (CIS) spectra of $E_{\rm B}\le 4.8$\,eV 
are shown in Fig.\,\ref{fig3}(c). 
One can see that the CIS spectra of the IGSs 
($E_{\rm B}\le 3.2$\,eV) showed 
Fano profiles \cite{Fano}: 
a local maximum was reached at $h\nu\sim 533.1$\,eV 
with a preceding dip at $h\nu\sim 531.6$\,eV 
[see also the intensity modulation at $E_{\rm B}<3$\,eV 
in Fig.\,\ref{fig3}(b)]. 
On the other hand, the line shapes of the 
CIS spectra of the O 2$p$ band region were similar to 
that of the absorption spectrum.

\begin{figure}[htb]
\begin{center}
\includegraphics[width=4.1cm]{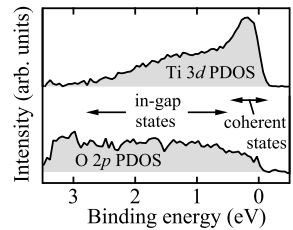}
\caption{\label{fig4} Ti 3$d$ and O 2$p$ PDOSs of STO:Nb near $E_{\rm F}$.}
\end{center}
\end{figure}

In Fig.\,\ref{fig4}, we show Ti 3$d$ and O 2$p$ PDOSs 
in the near-$E_{\rm F}$ regions derived 
from the resonant PES measurements. 
Here, the Ti 3$d$ (O 2$p$) PDOS is the difference 
between the spectra taken at 
$h\nu =$ 458.2\,eV (533.1\,eV) and 450.4\,eV (531.6\,eV.) 
One can see that the IGSs consist of 
Ti 3$d$ and O 2$p$ states, whereas 
the coherent states mainly consist of Ti 3$d$ states. 
The presence of Ti 3$d$ character in the IGS 
shows that this structure cannot solely be attributed to 
Nb 4$d$ donor levels.

If the IGSs of STO:Nb were formed by a spectral-weight 
transfer from the coherent states as schematically shown in 
Fig.\,\ref{fig0}(b), 
one would expect similar line shapes for the Ti 3$d$ and O 2$p$ PDOSs, 
which is apparently not the case as seen in Fig.\,\ref{fig4}. 
Therefore, one needs to explicitly consider the O 2$p$ states 
as well as the Ti 3$d$ states in explaining the coherent states 
and the IGSs. 
To this end, we adopt a local description of the electronic structure: 
the coherent states and IGSs originate from Ti 
sites initially having nominally-$d^1$ configuration 
(the same initial state to that of the Ti$^{3+}$ peak 
in the Ti 2$p$$_{3/2}$ spectrum). 
This interpretation is supported by the observation 
\cite{Robey, Yoshida_EPL} that the spectral weight of the 
region including the coherent states and the IGSs scales with $x$, 
which is similar to the Ti$^{3+}$-component weight in the 
Ti 2$p$$_{3/2}$ spectra scaling with $x$ \cite{Morikawa}. 
Then, the mixed character of $p$ and $d$ in the 
IGSs can be explained as 
the final state composed of $d^1\underline{L}$ 
($\underline{L}$ denotes a hole in the O 2$p$ ligand orbital) 
and $d^0$, since the former has 
$p$ character while the latter has $d$ character. 
The mixing of $d^1\underline{L}$ into $d^0$, which is a consequence of 
$d$-$p$ hybridization, can alternatively be described that the 
Ti 3$d$-electron-emitted final state is partly screened 
by the electrons in the 
local ligand (O 2$p$) orbitals. 
In fact, there is a recent trend to understand the valence-band spectra 
in terms of screening orbitals \cite{Okada_CT, Okada2, Mossanek_CSVO}, 
a concept originally developed to understand the core-level 
spectra \cite{Fuggle}. 
The coherent state, on the other hand, 
is understood as a final state which is 
non-locally screened by a coherent band having mainly-$d$ character 
\cite{Mossanek_CSVO, Taguchi}. 
These assignments are 
similar to those adopted for LaTiO$_3$ \cite{Okada2} and 
(Ca,Sr)VO$_3$ \cite{Mossanek_CSVO}, although STO:Nb is a non-integer filling 
system. 

Our assignment of the IGS to 
locally-screened 
incoherent state implies that there exists 
another incoherent state at higher $E_{\rm B}$ having poorly-screened 
character [Fig.\,\ref{fig0}(c).] 
Thus, the spectral weight of the coherent state of 
doped STO is transferred with increasing correlation ($x\to$ 1) 
to the poorly-screened incoherent state as well as to the locally-screened 
state [see the electronic-structure evolution with $x$ illustrated in 
Fig.\,\ref{fig0}(c).] This picture smoothly connects to the 
theoretical prediction \cite{Okada_CT} that the 
charge gap of the $d^1$ end member such as 
LaTiO$_3$ has intermediate character 
between Mott-Hubbard and charge-transfer-type 
due to strong $d$-$p$ hybridization 
(see also \cite{JHPark, Mossanek_CSVO}.)

It is well known that locally-screened final states, split off from the 
itinerant O 2$p$ band, are similar 
to multiple charge states of transition-metal impurities 
in semiconductors as proposed by Haldane and Anderson 
\cite{HaldaneAnderson, NaCuO2_Mizokawa, VanVeenendaal}. 
The present IGSs having locally-screened 
character are thus considered qualitatively 
equivalent to the multiple charge states. 
Here, the ``impurity" site is the Ti site having a nominally-$d^1$ 
configuration embedded in the host STO. 
With increasing the hybridization between the impurity and the host, 
multiple charge states of the impurity accumulate in the middle of 
the band gap since the energy separation of different charge states becomes 
small due to increased screening, and also since the states are more 
strongly repelled form the valence band and the conduction band 
\cite{HaldaneAnderson}. 
Hence, the IGS of electron-doped 
STO appearing in the middle of the band gap is understood 
as a result of strong $d$-$p$ hybridization. 
We expect that a similar mechanism could be applicable to 
explain e.g.\,the in-gap states of electron-doped TiO$_2$ and 
the incoherent states positioned $\sim$1\,eV below $E_{\rm F}$ of 
the moderately correlated vanadates 
\cite{Inoue_PRL, Sekiyama, Maiti_EPL, Eguchi_PRL, Mossanek_CSVO}. 
Direct investigation of O 2$p$ states as demonstrated here 
or the search for poorly-screened 
final states 
will be the keys to prove this senario.

We thank M.~Takizawa, A.~Fujimori, T.~Yoshida and D.-Y.~Cho 
for useful information and discussion.

\end{document}